\documentstyle[12pt]{article}
\begin{document}
\hsize37truepc\vsize61truepc
\hoffset=-.5truein\voffset=-1.3truein
\setlength{\baselineskip}{17pt plus 1pt minus 1pt}
\setlength{\textheight}{22.5cm}
\vphantom{0}\vskip1.1truein
\leftline{STRINGS and CAUSALITY}
\vskip.65truein

\hbox{\obeylines\baselineskip12pt\parskip0pt\parindent0pt\hskip1.1truein
\vbox{
\vskip.1truein
\leftline{\it Emil Martinec}
\smallskip
\leftline{Enrico Fermi Inst. and Dept. of Physics}
\leftline{University of Chicago, Chicago, IL 60637}
}}
\vskip .65truein
\noindent
%
%
\vskip.65truein

\def\be{\begin{equation}}
\def\ee{\end{equation}}
\def\bea{\begin{eqnarray}}
\def\eea{\end{eqnarray}}

\def\np{Nucl. Phys. }
\def\pl{Phys. Lett. }
\def\pr{Phys. Rev. }
\def\prl{Phys. Rev. Lett. }
\def\cmp{Comm. Math. Phys. }
\def\grg{Gen. Rel. and Grav.}
\def\cqg{Class. Quant. Grav.}
\def\journal#1&#2(#3){\unskip, \sl #1\ \bf #2 \rm(19#3) }
\def\andjournal#1&#2(#3){\sl #1~\bf #2 \rm (19#3) }
\def\nextline{\hfil\break}
\def\ie{{\it i.e.}}
\def\eg{{\it e.g.}}
\def\cf{{\it c.f.}}
\def\inbar{\,\vrule height1.5ex width.4pt depth0pt}
\def\IB{\relax{\rm I\kern-.18em B}}
\def\IC{\relax\hbox{$\inbar\kern-.3em{\rm C}$}}
\def\ID{\relax{\rm I\kern-.18em D}}
\def\IE{\relax{\rm I\kern-.18em E}}
\def\IF{\relax{\rm I\kern-.18em F}}
\def\IG{\relax\hbox{$\inbar\kern-.3em{\rm G}$}}
\def\IH{\relax{\rm I\kern-.18em H}}
\def\ii{\relax{\rm I\kern-.18em I}}
\def\IK{\relax{\rm I\kern-.18em K}}
\def\IL{\relax{\rm I\kern-.18em L}}
\def\IM{\relax{\rm I\kern-.18em M}}
\def\IN{\relax{\rm I\kern-.18em N}}
\def\IO{\relax\hbox{$\inbar\kern-.3em{\rm O}$}}
\def\IP{\relax{\rm I\kern-.18em P}}
\def\IQ{\relax\hbox{$\inbar\kern-.3em{\rm Q}$}}
\def\IR{\relax{\rm I\kern-.18em R}}
\font\cmss=cmss10 \font\cmsss=cmss10 at 7pt
\def\IZ{\relax\ifmmode\mathchoice
{\hbox{\cmss Z\kern-.4em Z}}{\hbox{\cmss Z\kern-.4em Z}}
{\lower.9pt\hbox{\cmsss Z\kern-.4em Z}}
{\lower1.2pt\hbox{\cmsss Z\kern-.4em Z}}\else{\cmss Z\kern-.4em Z}\fi}
\def\IGa{\relax\hbox{${\rm I}\kern-.18em\Gamma$}}
\def\IPi{\relax\hbox{${\rm I}\kern-.18em\Pi$}}
\def\ITh{\relax\hbox{$\inbar\kern-.3em\Theta$}}
\def\IOm{\relax\hbox{$\inbar\kern-3.00pt\Omega$}}
\def\sdtimes{\mathbin{\hbox{\hskip2pt\vrule height 4.1pt depth -.3pt width
.25pt \hskip-2pt$\times$}}}
\newdimen\xraise\newcount\nraise
\def\xpoint{\hbox{\vrule height .45pt width .45pt}}
\def\udiag#1{\vcenter{\hbox{\hskip.05pt\nraise=0\xraise=0pt
\loop\ifnum\nraise<#1\hskip-.05pt\raise\xraise\xpoint
\advance\nraise by 1\advance\xraise by .4pt\repeat}}}
\def\ddiag#1{\vcenter{\hbox{\hskip.05pt\nraise=0\xraise=0pt
\loop\ifnum\nraise<#1\hskip-.05pt\raise\xraise\xpoint
\advance\nraise by 1\advance\xraise by -.4pt\repeat}}}
\def\bdiamond#1#2#3#4{\raise1pt\hbox{$\scriptstyle#2$}
\,\vcenter{\vbox{\baselineskip12pt
\lineskip1pt\lineskiplimit0pt\hbox{\hskip10pt$\scriptstyle#3$}
\hbox{$\udiag{30}\ddiag{30}$}\vskip-1pt\hbox{$\ddiag{30}\udiag{30}$}
\hbox{\hskip10pt$\scriptstyle#1$}}}\,\raise1pt\hbox{$\scriptstyle#4$}}
\def\p {\partial}
\def\frac#1#2{{#1\over#2}}
\def\coeff#1#2{{\textstyle{#1\over #2}}}
\def\hf{\coeff12}
\def\half{\frac12}
\def\sst{\scriptscriptstyle}
\def\txt{\textstyle}
\def\bra#1{\left\langle #1\right|}
\def\ket#1{\left| #1\right\rangle}
\def\vev#1{\left\langle #1 \right\rangle}
\def\det{{\rm det}}
\def\tr{{\rm tr}}
\def\mod{{\rm mod}}
\def \sinh{{\rm sinh}}
\def \cosh{{\rm cosh}}
\def \sgn{{\rm sgn}}
\def\det{{\rm det}}
\def\exp{{\rm exp}}
\def\sh{{\rm sinh}}
\def\ch{{\rm cosh}}
\def\One{{1\hskip -3pt {\rm l}}}
\catcode`\@=11
\def\slash#1{\mathord{\mathpalette\c@ncel{#1}}}
\overfullrule=0pt
\def\steepslash{\c@ncel}
\def\frac#1#2{{#1\over #2}}
\def\AA{{\cal A}}
\def\BB{{\cal B}}
\def\CC{{\cal C}}
\def\DD{{\cal D}}
\def\EE{{\cal E}}
\def\FF{{\cal F}}
\def\GG{{\cal G}}
\def\HH{{\cal H}}
\def\II{{\cal I}}
\def\JJ{{\cal J}}
\def\KK{{\cal K}}
\def\LL{{\cal L}}
\def\MM{{\cal M}}
\def\NN{{\cal N}}
\def\OO{{\cal O}}
\def\PP{{\cal P}}
\def\QQ{{\cal Q}}
\def\RR{{\cal R}}
\def\SS{{\cal S}}
\def\TT{{\cal T}}
\def\UU{{\cal U}}
\def\VV{{\cal V}}
\def\WW{{\cal W}}
\def\XX{{\cal X}}
\def\YY{{\cal Y}}
\def\ZZ{{\cal Z}}
\def\C{{\bf C}}
\def\K{{\bf K}}
\def\Z{{\bf Z}}
\def\lam{\lambda}
\def\eps{\epsilon}
\def\vareps{\varepsilon}
\def\del{\delta}
\def\hb {\hbar}
\def\SL{${\rm SL}_2$}
\def\brs{{\scriptscriptstyle\rm BRS}}
\def\zbar{\bar z}
\def\ddz{\partial_z}
\def\ddzbar{\partial_{\bar z}}
\catcode`\@=12
%
\def\ppl{{p^+}}
\def\pmi{{p^-}}
\def\pperp{{\vec p}}
\def\xpl{{x^+}}
\def\xmi{{x^-}}
\def\xmiz{{x^-_0}}
\def\xperp{{\vec x}}
\def\xnperp{{{\vec x}_n}}
\def\Xpl{{X^+}}
\def\Xmi{{X^-}}
\def\Xperp{{\vec X}}
\def\pz{{p^0}}
\def\A{{\bf A}}
\def\aa{{\bf a}}
\def\rgyr{R_{\rm gyr}}

One of the early concerns in quantum field theory
was causality \index{causality}--
can one ensure that measurements at spacelike separation do not interfere?
General issues of this type were important in the days when little
was understood about the underlying dynamics of particle interactions;
the hope was to place constraints on the types of allowed
dynamics and interactions on the basis of cherished tenets of kinematics.
There were in fact two ways to approach the problem.  First, one could
take the point of view that only the results of scattering experiments
had physical content; then causal behavior means that the scattered
wave cannot reach the detector before the incident wave strikes the
target.  We might describe this as global causality, and it is
obeyed by tachyon-free string theory \index{string theory}
S-matrices.  However there is a
second, local version of causality,
which asks whether local measurements commute at spacelike separation.
Local causality ensures global causality in ordinary field theory,
and typically one can construct global violations from local ones,
so it would seem that the two are equivalent.
But in quantum field theory the local question is more fundamental,
and can be resolved without solving any complicated global problem
such as long-distance signal propagation.
Local causality rests on the universal properties
of field theory, \eg\ that any manifold locally looks the
same, rather than their implementation in any particular
spacetime background.  For instance we can be sure that an
interacting  scalar
field propagating in a Schwarzschild geometry obeys the axioms of
local field theory without having to solve for the full S-matrix
of the problem.

The consequences of causality in field theory
are dispersion relations arising from the analyticity of
correlation functions in the complex plane of the kinematical invariants.
These dispersion relations reaveal themselves in terms of the
analyticity properties of the correlation functions\cite{kallen}.
For instance the $N$-point correlation function of a scalar field
\bea
  \langle 0|\phi(x_1)\cdots \phi(x_N)|0\rangle =
	\int ds_{ij}\; G_N(s_{ij})\; \Delta_N(x_i-x_{i+1};s_{ij})
	\label{grfn}\\
  \Delta_N =\int dp_i\;\exp[ip_j\cdot(x_j-x_{j+1})]
  	\delta(p_i\cdot p_j - s_{ij})\prod_k \theta(p_k^0)
	\label{deltan}
\eea
is analytic in the complex plane of the kinematical invariants $s_{ij}$
(for the two-point function this is the K\"allen-Lehmann
representation of the propagator).  The representation splits
dynamics contained in the function $G_N$ from the kinematics
residing in the basic objects $\Delta_N$.  Dispersion relations arise
from the deformation of the contour integral over $s_{ij}$
onto cuts.

In quantum particle-field theory, we are used to the idea that dynamics
takes place on a given spacetime manifold, with specified causal structure.
A generic kinematic structure provides the arena in which dynamics unfolds.
Causality of the dynamics is intimately tied to the
topology of light cones.
One of the great confusions of the present day is how to reconcile
fluctuations of the causal structure and causality; how can
quantum superposition and causal relationships of events coexist?
It is an issue the Hawking paradox frames sharply.

String theory is in an even more primitive
state.  I think it is fair to say that the issue of what a light cone
is in string theory is not yet settled even in the
classical theory.  The string S-matrix is causal when considered
as a function of the center of mass momenta of scattering strings.
However the center of mass momentum is conjugate to the center of
mass position of the string, which may or may not be causally related
to the locations of string interactions which take place locally
on the string; it is not a priori obvious how to relate this
global causality in the center of mass to some local notion
of causality in, say, the loop space that is the arena
of string dynamics as we now understand it.
There is of course
the natural notion that the light cone \index{light cone}
structure of loop space \index{loop space} is
induced from the light cone topology of its underlying point manifold.
But can there be a different notion of light cone?  Indeed,
is there any operational sense in which the light cone structure
is determined by the point manifold light cones when there are
no pointlike objects which could detect this causal structure?
For instance, with the infinity of points on the string, there are
infinitely many Lorentz invariant quantities one can build;
what could be the analog of $\Delta_N$ in string field theory?

A number of arguments have been put forward proposing
that string theory
exhibits a `generalized uncertainty principle'
whereby objects cannot be localized to a region smaller than
the string scale $\ell_s=\sqrt{\hbar c\alpha'}$
\cite{uncertainty}\cite{renormuncertainty}\cite{grossmende}
\cite{klebsuss}\cite{atickwitten}
There are two
models for how this principle manifests itself in string theory.
The first model is entropic; it is not that there is a limiting size
in string theory, but rather that the {\it probability}
of observing pointlike behavior is extremely small (\eg\
form factors for hard scattering have gaussian falloff as
one begins to probe constituent structure\cite{grossmende}, or
thermal partition functions exhibit slower growth at
high temperature\cite{atickwitten}).
Such a model would not require a fundamental overhaul of
the conceptual foundations of spacetime geometry.
The second model is the Heisenberg uncertainty relation --
short distance measurements are meaningless and we should reformulate
the theory in such a way that it is impossible to ask about them.
The irrelevance of spacetime discreteness\cite{klebsuss}\cite{ccc}
and the smearing introduced by renormalization\cite{renormuncertainty}
may point in this direction.  If this second model
were the proper setting, then one would probably need to abandon the notion
of having the string causal structure induced from a point manifold,
if not the notion of manifolds and loop space altogether.
Obviously the choice of model made by string theory
has profound implications for the question of what a horizon
(and hence a black hole) is in string theory.

One approach to the causality problem is to work backwards: to take a given
string theory and try to deduce the causal structure from a specific
set of measurements.  This has the unfortunate property that
it does not necessarily isolate the kinematics from the dynamics.
In addition, the specific measurements we shall choose below are not
gauge invariant, and eventually we will have to understand how to separate
the causal structure and the gauge dependence.
In particle-field theory, the light cone is quite simply found
as the boundary of the domain of commutativity of two scalar field
operators.  We will adopt the same definition for string theory
and see where it leads us.

Consider the two-point function in particle field theory
\be
  \bra{0}[\phi(x),\phi(x')]\ket{0} =
        \oint_\CC \frac{dp^0}{2\pi}\int\frac{d\pperp}{(2\pi)^{d-1}}
        \ \frac{i\;e^{ip\cdot x}}{p^2-m^2}\ .
\label{comm}
\ee
The contour integral here encircles both the poles of the integrand
in the complex $p^0$ plane.  Some elementary algebra yields
\be
  \bra{0}[\phi(x),\phi(x')]\ket{0} =
        \int_0^\infty d\tau \Bigl(\frac{1}{4\pi\tau}\Bigr)^{d/2}
        \biggl(\exp\;i\biggl[\frac{(x-x')^2}{4\tau}-m^2\tau\biggr]
        \quad - \quad h.c.\biggr)\ .
\label{commie}
\ee
This expression has the following analyticity properties
in $\tau$ as a function of $(x-x')^2$:
\begin{itemize}
\item $(x-x')^2>0$:  Both terms in (\ref{commie}) are real and equal
upon continuation to the approporiate imaginary semi-axis;
they cancel one another and the field commutator vanishes.
\item $(x-x')^2<0$:  One cannot rotate the contour to the imaginary
axis so that the integral converges both at zero and infinity;
the two terms don't cancel and the field commutator is nonzero.
\end{itemize}
Thus the hypersurface $(x-x')^2=0$ can be identified as the
boundary of causal propagation -- the light cone.

The signs of the exponent for $\tau\rightarrow 0$ and
$\tau\rightarrow\infty$ determine the
direction of the contour rotation.  The sign for $\tau\rightarrow
\infty$ is controlled by the sign of the particle mass.
The sign for $\tau\rightarrow 0$ can be read directly
from the semiclassical limit of the path integral representation
of the two-point function
\be
  \bra{0}\phi(x)\phi(x')\ket{0}=\int_{X(0)=x\atop X(\tau)=x'} \DD X
	\;\exp\left[\coeff i{m\hbar}\int_0^\tau \coeff12{\dot x}^2\right]
\label{twopt}
\ee
For short proper time of propagation,
the straight-line motion of the particle from initial to final
points dominates: $S_{{\rm cl}}\propto(x-x')^2/\tau$.
The sign of the exponent is indeed the sign of $(x-x')^2$.

The calculation in string field theory (as currently understood)
is no different.  The string field $\Phi(X(\sigma),B(\sigma),C(\sigma))$
creates/destroys entire strings
(here $B(\sigma)$, $C(\sigma)$ are Faddeev-Popov ghost
coordinates).
The two-point function
can be written as a path integral\cite{stringprop}\cite{cmnp}
with the action
\be
  \bra{0}\Phi(X)\Phi(X')\ket{0}=
	\int_{X_{{\rm init}}=X(\sigma)\atop X_{{\rm final}}=X'(\sigma)}
	\DD X
	\;\exp\left[\coeff i{\hbar\alpha'}\int_0^\tau
	dt\;d\sigma [-(\partial_t X)^2 +(\partial_\sigma X)^2]\right]
\label{strtwopt}
\ee
In order to simplify the presentation the ghost coordinates have been
suppressed here and below.
The short-time limit of the action ignores the harmonic forces
$(\partial_\sigma X)^2$ since there is no time to react to them.
Each point on the string moves ballistically from $X(\sigma)$
to $X'(\sigma)$, so the classical action is
\be
S_{{\rm cl}}\sim\frac{\int d\sigma (X-X')^2}{\tau}
\label{scl}
\ee
Hence repeating the steps to the point particle field commutator yields
$[\Phi,\Phi']=0$ if and only if $\int d\sigma(X-X')^2>0$.
Note that there is an immediate generalization of the
causal boundary to an arbitrary curved spacetime
\be
\int d\sigma\; \II(X(\sigma),X'(\sigma))=0
\label{curved}
\ee
where $\II(\sigma)$ is the geodesic interval between $X(\sigma)$
and $X'(\sigma)$.

Let me conclude with a few remarks:
\begin{itemize}
\item Any given measurement can only reveal
a finite amount of information about the relative location of the
two strings.  So the strings must decide whether they are relatively
spacelike or timelike -- yes or no -- and faced with only
these two options, {\it democratically takes the majority vote}
across the string.  But note that this answer is quite bizarre since
a given measurement can only reflect the average causal relationship
of the points on the two strings.
Consider the case where $X(\sigma)$ is a pointlike string, $X(\sigma)=x$.
Then part of $X'(\sigma)$ can be {\it outside} the
point light cone of $x$, yet still the measurements interfere;
on the other hand, part of $X'(\sigma)$ can be {\it inside}
the point light cone of $x$ and yet the measurements
do not interfere!
The fact that the commutator calculation can be phrased in terms of
semiclassical world sheet physics does mean that no information encoded
in the string configuration itself is being propagated outside the point
light cone; the Virasoro constraints guarantee that the
world sheet causal structure is the one projected from the point
spacetime.  However the string field is a function on the whole string,
not part of it, so it is not clear what the effect is on
the propagation of the string field.
\item The light cone
\be
\int d\sigma(X-X')^2=0
\label{lightcone}
\ee
is invariant under conformal transformations of spacetime generated
on loop space by the operator
$\int d\sigma X(\sigma)\partial/\partial X(\sigma)$
and its cousins,
so it does play to some extent the same role that the point light
cone does for a point particle.
\item Unfortunately the expression (\ref{lightcone})
is rather badly noninvariant
under gauge transformations of string theory (reparametrizations
of the loops).  Thus one might worry that `spacelike' noncommutation of
string fields is only a gauge artifact; however the fact that a light
cone calculation yields a similar result \cite{martinec}
suggests that this is not the crux of the problem.
\item The semiclassical approach outlined above immediately shows that
the evaluation of field commutators in the interacting theory yields
results that are strongly dependent on the choice of interaction
vertex.  For instance, the correlation function
\be
\bra{0}\Phi(X_1)[\Phi(X_2),\Phi(X_3)]\ket{0}
\label{threepoint}
\ee
can be evaluated with either the Witten-type\cite{witten}
or Mandelstam-type\cite{mandelstam}
overlap.  The sign of the exponent in the short-time, semiclassical limit
of the path integral will give a vanishing result a) for
the Witten vertex if the
overlapping `half-strings' obey condition $\int_a^b (X_2-X_3)^2>0$ (with
$a$ the string midpoint and $b$ its endpoint),
and b) for the Mandelstam vertex if the segments of strings two and three
that overlap (\eg\ all of string two overlaps with part of
string three) are on average causally related.
In either case the answer seems not to depend on the parts of
the loop arguments outside the segments that overlap, and does
depend strongly on the (gauge-dependent) geometry of the overlap.
All this points to the necessity of a better understanding of
string gauge invariance before causality can be properly addressed.
\item We have phrased the problem of causality in the loop representation
(position eigenstates $X(\sigma)$).  Causality for mass eigenstates
involves a convolution
\be
  \Phi(\{n_\ell\})=\int \prod_\ell dx_\ell\;
	H_{n_\ell}(x_\ell)\Phi(\{x_\ell\})
\label{convolve}
\ee
where $H_{n_\ell}(x_\ell)$ is a Hermite function.
This smearing involves $\delta x_\ell\sim 1/\ell$; summing over $\ell$,
the string wanders logarithmically over all spacetime.  The
typical string contributing to typical processes is wild;
smooth loops have measure zero.  But then
what is causality?  Clearly we need a renormalized notion of
the spread of the string and of locality.  In the low energy theory,
particles are a good approximation to strings; only the center of
mass location of the string is effectively measured, and the
logarithmically large spread of the string is erased by
averaging to leave a finite but nonzero residue of order the string scale.
The question is whether in this averaging process
one smears out the `location' of information carried by strings.
An effective string size should be set by the spacetime energy scale
of the process under consideration (as occurs, for example,
in \cite{grossmende}).  This sort of consideration
could result in an effective smearing of the light cone.
Similar arguments have been put forward\cite{uncertainty}
to support the Heisenberg model of the generalized uncertainty principle.
\end{itemize}

Finally, and perhaps most important,
the argument leading to the causality condition assumed that
the fluctuation determinants of the string modes do not compete
with the exponential of the classical action in determining the
convergence of the proper time integral in the path integral representation
of the two-point function.
Indeed, in particle-field theory the determinant
contributes a power law underneath $exp[-S_{cl}]$.
However we should know very well that in string theory this
is usually not the case due to the exponential
growth in the level density of string states;
I thank M. Green for pointing this out to me.
The effect of including this exponential contribution is to shift
the commutator condition by a constant (spacelike) term (see \cite{green}
for closely related calculations):
\be
\int d\sigma(X-X')^2=1
\ee
is the causal boundary.  Green has suggested that this is another
piece of evidence pointing toward the Heisenberg paradigm for the
generalized uncertainty relation, since even pointlike states
leak information outside the point light cone\cite{cmnp}.
This spacelike shift appears even in the superstring for certain
worldsheet boundary conditions and hence seems unrelated to the
presence or absence of a tachyon in the physical spectrum.
If these boundary conditions contribute to physical processes,
and the spacelike pole is not a gauge artifact\cite{cmnp},
then indeed we will have to revise our notions of geometry
in string theory.

\vskip .5in
{\bf Acknowledgements}: I want to thank Michael Green for a stimulating
discussion and an explanation of his work.  This work was supported
in part by Dept. of Energy grant DEFG02-90ER-40560.

\end{document}